\documentclass[twocolumn,aps,nobibnotes,prl,superscriptaddress,amsmath,amssymb,amsfonts]{revtex4-1}
\usepackage{graphicx}
\usepackage{color}
\usepackage{ulem}
\usepackage{url}
\usepackage{multirow} 
\usepackage{dcolumn}

\def\bea{\begin{eqnarray}}
\def\eea{\end{eqnarray}}
\def\ben{\begin{equation}}
\def\een{\end{equation}}
\def\benu{\begin{enumerate}}
\def\enu{\end{enumerate}}
\def\bei{\begin{itemize}}
\def\eei{\end{itemize}}

\def\sss{\scriptscriptstyle\rm}

\def\1var{(\bx_1...\bx\N)}

\def\br{{\bf r}}

\def\bx{{\br t}}

\def\N{_{\sss N}}

\def\sph_int{ {\int d^3 r}}

\begin{document}

\normalem 

\title{Ultrafast Local Magnetization and Demagnetization in Heusler Alloys}
\author{P. Elliott}
\affiliation{Max-Planck-Institut f\"ur Mikrostrukturphysik, Weinberg 2, D-06120 Halle, Germany.}
\author{T. M{\"u}ller}
\affiliation{Max-Planck-Institut f\"ur Mikrostrukturphysik, Weinberg 2, D-06120 Halle, Germany.}
\author{J. K. Dewhurst}
\affiliation{Max-Planck-Institut f\"ur Mikrostrukturphysik, Weinberg 2, D-06120 Halle, Germany.}
\author{S. Sharma}
\affiliation{Max-Planck-Institut f\"ur Mikrostrukturphysik, Weinberg 2, D-06120 Halle, Germany.}
\affiliation{Department of physics, Indian Institute for technology-Roorkee, 247667 Uttarkhand, India}
\author{E. K. U. Gross}
\affiliation{Max-Planck-Institut f\"ur Mikrostrukturphysik, Weinberg 2, D-06120 Halle, Germany.}

\date{\today}

\begin{abstract}
With the goal of pushing Spintronic devices towards faster and faster timescales, we demonstrate, using {\it ab-intio} time-dependent density functional theory simulations of bulk Heusler compounds subject to ultrashort intense laser pulses, that the local magnetic moment can increase or decrease in a few femtoseconds. This speed is due to the all optical 
nature of the process, which transfers spin moment from one sublattice to another. This transfer depends on easily tunable laser parameters. 
By comparing the spin dynamics of a variety of Heusler (or half-Heusler) compounds, we demonstrate that the density of states explains the observed moment transfer; most the physics of inter sublattice moment transfer is due to the flow of spin current which is governed by availability of states above the Fermi level. 

\end{abstract}

\maketitle


Femtomagnetism\cite{U09,ZHBB02}, whereby electronic spins are manipulated at femtosecond timescales, opens up a whole new field for ultrafast devices which are several orders of magnitude faster than those currently available\cite{TSKK04}. The fastest timescale on which devices operate is dictated by the internal processes by which the spins can be modified. To fully exploit the possibilities of controlling these ultrafast processes, it is crucial to know  the limitations on the timescale on which this manipulation of spins is possible. Recently there has been much effort on making these timescales faster, e.g. spin-reorientation in anti-ferromagnets was found to be much faster than previously anticipated\cite{KKTP04,KKR10}.

The most advantageous processes in terms of speed and ease of manipulation are those that take place in the coherent regime, i.e. dissipative phenomena such as coupling to electron-phonon scattering, are not yet dominant. In this regime it is purely electronic processes that control the physics and since electrons couple directly to the laser, it allows one to benefit from the extraordinary progress in creating ultrashort laser pulses over the past few decades.
It is in this regime we will focus our attention.

Currently, experiments\cite{BMDB96,HMKB97,SBJE97,ABPW97,RVK00,KKKJ00,KULM02,SPWD05,SKPM07,MRWP11,EBMP13,TLSG13} investigating laser-induced spin dynamics all observe a loss of magnetic moment at short timescales. However, for full controllability of the moment, a process by which the moment is increased is required. We show that in the case of materials with several magnetic sublattices, it is possible to see an increase in the magnetic moment before the demagnetization sets in. 
Although, this behavior is observed for a particular class of materials, namely the Heusler family\cite{GFP16}, we demonstrate that the physics of this phenomena is  controlled by the ground-state density-of-states (DOS). This link between the observed magnetization dynamics and the easily calculable ground-state DOS makes tailoring of materials for ultrafast spin manipulation a clear possibility. Furthermore, we show this interpretation is valid for both ferromagnetically and antiferromagnetically coupled materials, further expanding the realm of potential materials which can exploit this form of spin dynamics.

\begin{figure}[th]
\centerline{\includegraphics[trim={0 0 1cm 0},clip,width=\columnwidth,angle=-0]{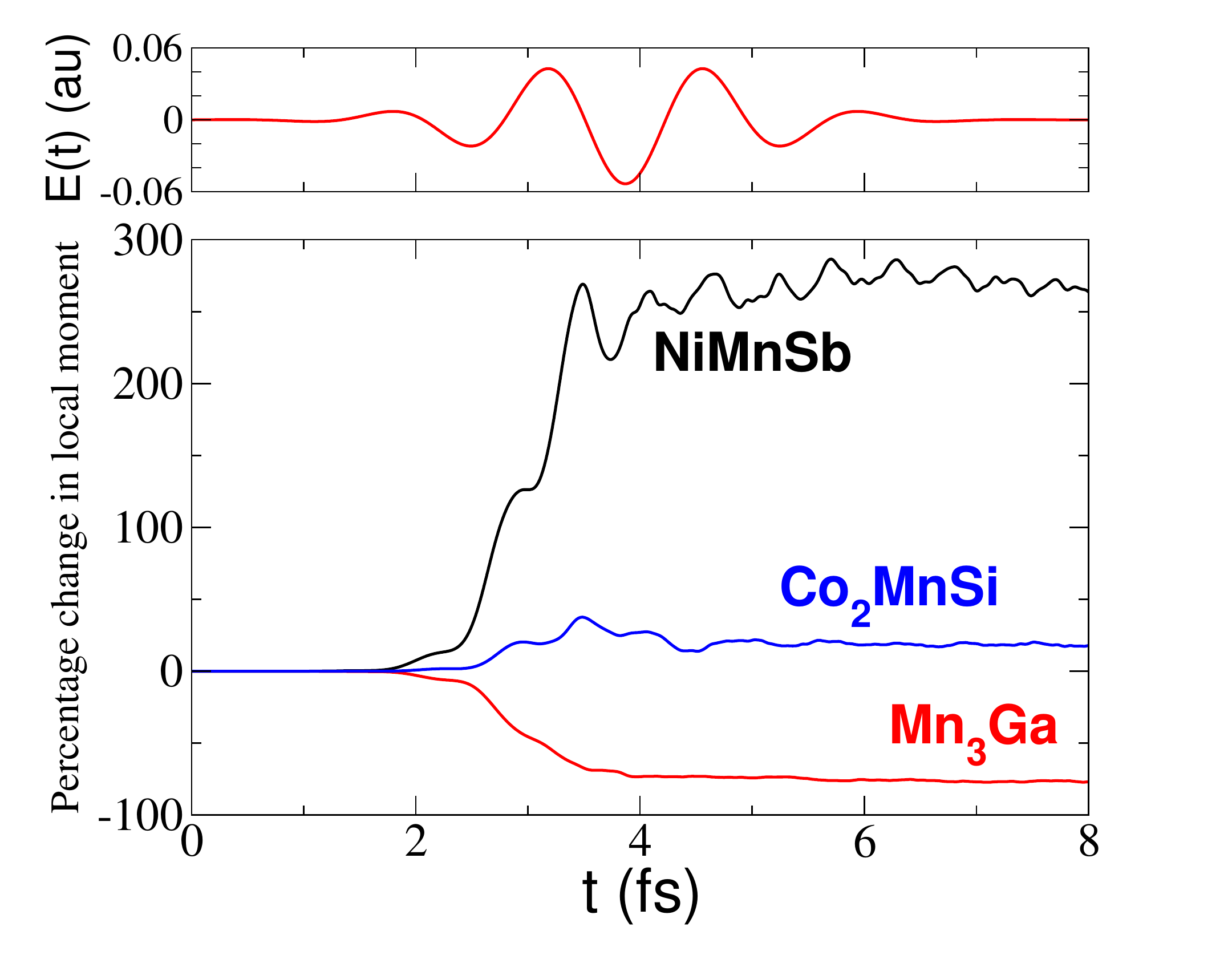}}
\caption{Upper panel: The electric field, $E(t)$, of the applied laser pulse. Lower panel: The relative change in the local magnetic moment on the X position atom for Heusler compounds NiMnSb, Mn$_3$Ga, and Co$_2$MnSi, following the Heusler stoichiometry X$_2$YZ (or XYZ for half-Heusler compounds).}
\label{f:perX8fs}
\end{figure}

The Heusler compounds we will study are Co$_2$MnSi, NiMnSb, and Mn$_3$Ga with the structural and ground-state details given in Table \ref{t:lat}. This choice was motivated by the fact that despite belonging to the same class they display very different magnetic behavior, thereby allowing us to explore a wide spectrum of possibilities-- all these materials have two magnetic sublattices which are anti-ferromagnetically coupled in Mn$_3$Ga and ferromagnetically coupled in the other two systems. NiMnSb shows a strong difference in the magnetic moment between the two sublattices, while this difference is not so pronounced in Co$_2$MnSi (see table \ref{t:lat} for details).

To study the spin and charge dynamics in these materials under the influence of ultrafast laser pulses, we will utilize the {\it ab-initio} method of time-dependent density functional theory (TDDFT). For more details on TDDFT see Refs. \cite{EFB09,TDDFTbook12,C11}, and in particular for its application to spin dynamics see Refs. \cite{KDES15,EKDS16}. To perform our calculations we will use the full potential linearized augmented-plane-wave method with $2$-component spinors, as implemented in the ELK\cite{elk} code. In all calculations a regular mesh in {\bf k}-space of $8\times8\times8$ grid points was used and a timestep of $\Delta t=0.05$ au used for the time-propagation algorithm\cite{KDES15}. The laser field applied in all cases is shown in the upper panel of Fig. \ref{f:perX8fs}, the pulse parameters are: frequency $\omega=2.72$ eV, FWHM=$2.42$ fs, and a fluence of $93.5$ mJ/cm$^2$ giving a peak intensity of $1\times10^{14}$ W/cm$^2$. The purpose of this pulse is to disentangle the optical excitation process from any subsequent dynamics such as spin-orbit mediated demagnetization. Later we will also perform calculations with much weaker (down to $1\times10^{11}$ W/cm$^2$) pulses.

\squeezetable
\begin{table}[tbh]
\caption{Relevant structural and ground-state magnetic properties of the Heusler compounds investigated in this work. All calculations used the LDA exchange-correlation (XC) functional.}
\label{t:lat}
\begin{ruledtabular}
\begin{tabular}{ c r|| l d | l d | l d }
& & \multicolumn{2}{c|}{\bf NiMnSb} & \multicolumn{2}{c|}{\bf Co$_2$MnSi} & \multicolumn{2}{c}{\bf Mn$_3$Ga} \\
\hline
\hline
\multicolumn{2}{c||}{Structural phase} & \multicolumn{2}{c|}{C$1_b$} & \multicolumn{2}{c|}{L$2_1$} & \multicolumn{2}{c}{D$0_{22}$} \\
\hline
\multicolumn{2}{c||}{Lattice parameters} & \multicolumn{2}{c|}{$a=5.90$} & \multicolumn{2}{c|}{$a=5.64$} & \multicolumn{2}{c}{$a=3.77$} \\
\multicolumn{2}{c||}{(\AA)} & & & & & \multicolumn{2}{c}{$c=7.16$} \\
\hline
         		& X & Ni & +0.30 & Co & +1.05 & Mn(2,3) & +2.01 \\
Local Moments  	& Y & Mn & +3.62 & Mn & +2.91 & Mn(1)   & -2.46 \\
($\mu_B$)  		& Z & Sb & -0.05 & Si & -0.04 & Ga      & -0.02 \\
\hline
Total Moment/atom ($\mu_B$) & & \multicolumn{2}{c|}{$1.33$} & \multicolumn{2}{c|}{$1.25$} & \multicolumn{2}{c}{$0.41$}
\end{tabular}
\end{ruledtabular}
\end{table}

In Fig. \ref{f:perX8fs} we plot the percentage change in the local magnetic moment, as a function of time, for the X atom (see table \ref{t:lat}) in each Heusler compound, relative to their respective ground-state moment. This magnetization dynamics is caused by the applied external laser field shown in the upper panel of Fig.\ref{f:perX8fs}.  The three studied materials behave very differently; the moment changes dramatically for Ni in NiMnSb with a gain of over $300\%$, the two Mn(2,3) atoms of Mn$_3$Ga lose almost all their moment ($100\%$) and the moment on Co atom stays effectively unchanged ($18\%$ gain).
It is interesting to note that these changes take place on an even faster timescale than the spin-orbit mediated dynamics observed in Refs. \cite{KDES15,ZH00}. The spin dynamics in this case is purely due to optical excitations, and so the global moment remains unchanged, with moment either transferred from one magnetic sublattice to another or transferred to the high-lying delocalized states. Any experimental measurements that are atom-sensitive will observe these changes, and, in fact, such an increase in NiMnSb may have already been observed\cite{Steilthesis}.

\begin{figure}[th]
\centerline{\includegraphics[trim={0 0 8.5cm 0},clip,width=\columnwidth,angle=-0]{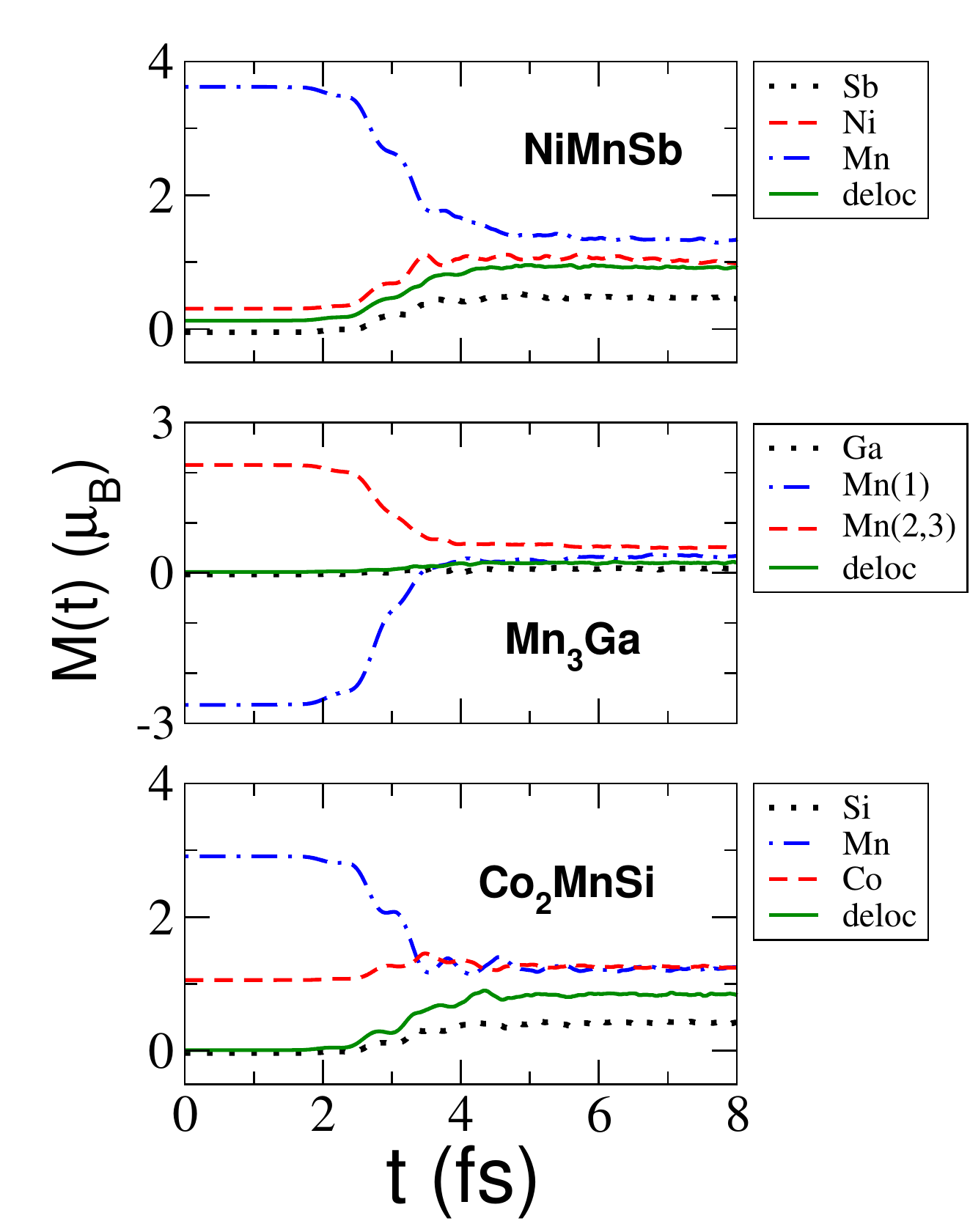}}
\caption{The dynamics of the local magnetic moment on each atom and of the delocalized moment for the Heusler compounds under consideration.}
\label{f:MT8fs}
\end{figure}

\begin{figure}[t]
\centerline{\includegraphics[width=\columnwidth,angle=-0]{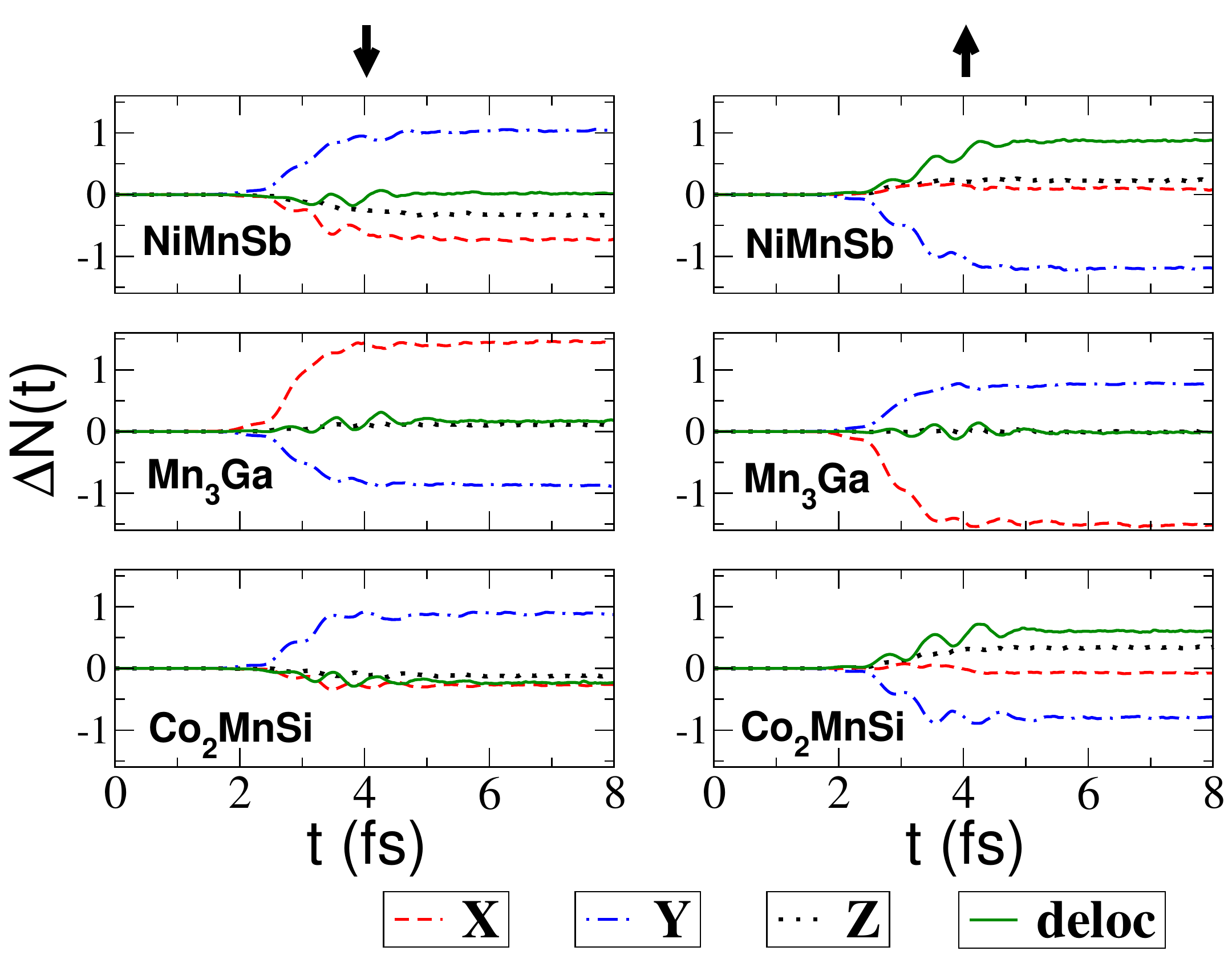}}
\caption{The time-dependent change in the spin-down (left) and spin-up (right) electrons on each atom (and of the delocalized electrons) relative to the ground-state for NiMnSb (upper panels) and Co$_2$FeSi (middle panels) and Mn$_3$Ga (lower panels). As in Fig. \ref{f:perX8fs}, the XYZ convention may be used to identify the respective atoms in each case. 
}
\label{f:NupNdw}
\end{figure}

For an in-depth understanding of this spin transfer process, we calculate the changes in the up and down spins  of each atom using:
\ben
\Delta N_{\uparrow\downarrow}(t) = \frac{\Delta N(t) \pm \Delta M_z(t)}{2}
\een
where $\Delta N(t) = N(t) - N_0$ is the change in the charge on the atom compared to the initial time, $N_0=N(t=0)$ and similarly for the moment $\Delta M_z(t) = M_z(t)-M_z(0)$. 
This definition is valid only when system is collinear. In the present case the system stays almost collinear despite the presence of spin-orbit coupling (SOC) and the external laser field, both of which allow for interatomic non-collinearity.

In Fig. \ref{f:NupNdw} we plot the change in the up and down spins on each atom for the three cases under investigation. In ferromagnetic materials with positive global moment, the
loss of up or gain of down electrons would lead to an increase in moment, while a gain of down or a loss of up spin electrons would lead to decrease in moment. For anti-ferromagnetic materials, the situation is more complex, if the local moment is positive then the same logic applies, but if the local moment is negative, a gain of up or a loss of down will cause a loss in moment, while a gain of down or a loss of up will lead to a gain of moment (i.e. an increase the magnitude of this moment). We first look at ferromagnetically coupled systems; in the case of NiMnSb, the $300\%$ increase in the moment of the Ni sublattice is due to the loss of down spin electrons by the Ni atom. These electrons are transferred to the Mn sublattice, leading to a decrease in the moment of the Mn sublattice. There is a further decrease in the moment on the Mn sublattice due to a transfer of some of the Mn spin-up electrons to the higher lying delocalized states. In the case of Co$_2$MnSi there is also a small amount of down spins transferred from Co sublattice to Mn sublattice, but this is compensated by a small transfer of up spins away from Co, leaving the local moment almost unchanged. 
In the case of anti-ferromagnetic Mn$_3$Ga down spins are transferred from the Mn(1) atom to the Mn(2,3) sublattice leading to a decrease in the moment of both sublattices. This decrease if further enhanced by up spins being transferred from the Mn(2,3) sublattice to the Mn(1) sublattice.

\begin{figure}[t]
\centerline{\includegraphics[width=\columnwidth,angle=-0]{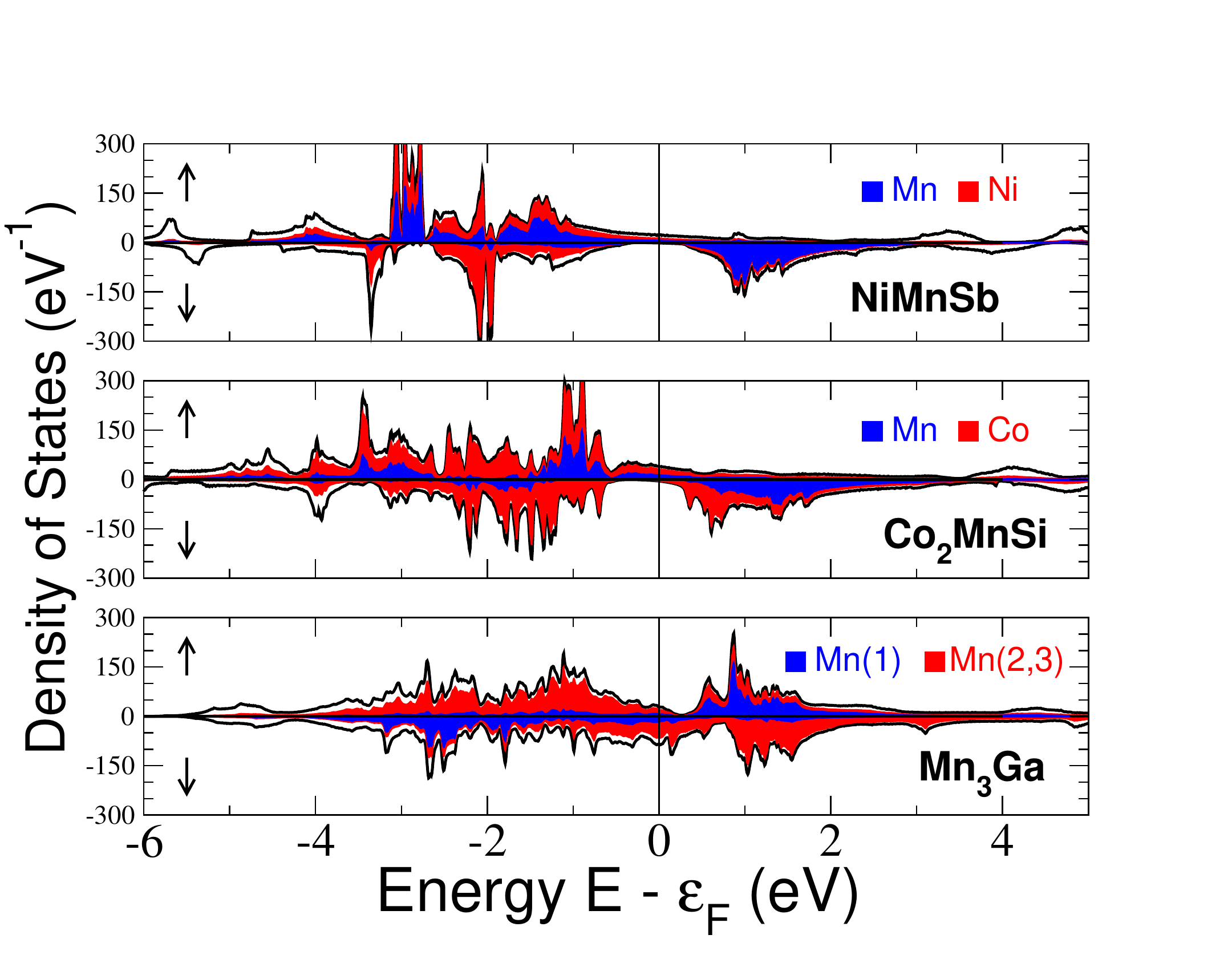}}
\caption{The total density of states for NiMnSb (upper panel), Co$_2$MnSi (middle panel), and Mn$_3$Ga (lower panel) around the Fermi level (set to zero) for the up and down electrons. In each case the contribution of the d-states localized to either the X and Y postion atoms to the total DOS is represented by the area filled.}
\label{f:DOS}
\end{figure}

Although, this transfer of up/down electrons provides an insight into the physics of gain/loss of moment, it does not tell us why this happens differently for the three materials studied and how one can tailor materials to obtain a desired spin-dynamics. We will now show that this spin transfer can be understood based on the ground-state density of states, which is an easily calculable quantity and can be used for material searches. After such a search is performed, the results can be verified by then performing TDDFT simulations. In Fig. \ref{f:DOS}, we show the total and partial DOS (site and $d$-projected) for all three materials. If we first look at the down spin DOS of NiMnSb, we can see that below the Fermi level the DOS is completely dominated by Ni states, whereas above, it is dominated by Mn states. Thus the optical excitation excites from the spin-down Ni states to the empty Mn $d$-states, causing spin-down current to flow from Ni sublattice to Mn sublattice. The Mn$_3$Ga spin transfer can be similarly explained by the DOS; for down spin, we see occupied Mn(1) states and unoccupied Mn(2,3) and vice versa for up spin, thus the laser will excite both these transitions, causing the moment to be lost in both sublattices. The lack of spin transfer from Co to Mn in Co$_2$MnSi can also be explained in a similar fashion;  below the Fermi level the spin-down DOS is dominated by Co $d$-states but, unlike for the case of NiMnSb and Mn$_3$Ga, there is an equal number of Co and Mn empty down states above the Fermi level. Due to available states on the same atom, it is more favorable for almost no spin current to flow from one sublattice to another. This then tells us that most of the physics of inter sublattice moment transfer is due to the flow of spin current which is governed by availability of states above the Fermi level.

We can further validate this DOS interpretation by performing additional calculations for different systems with similar DOS or with different laser parameters. Since the spin transfer process is based on pure optical excitations, it relies on the availability of states reachable with the applied laser pulse. Based on the energy difference between the occupied and empty states in the DOS, we can expect significant spin transfer due to excitations with an applied laser field with frequency $\sim 3$ eV. Choosing a laser frequency such that optical transitions are weak (e.g. $\omega=5.44$ eV), as expected, there is significantly less moment loss by the Mn atom in Co$_2$MnSi is observed.

Turning now to a different a set of materials, but with similar DOS; the compound Ni$_2$MnGa, whose DOS resembles NiMnSb, also shows the same qualitative behavior as NiMnSb (a gain of $190\%$ by the Ni atom). Similarly, the partial DOS of Co$_2$FeSi is almost the same to that of Co$_2$MnSi and, as expected, we find that the behavior of the local moments is also the same (a $7\%$ increase on the Co atoms). Note that for Co$_2$FeSi it is known that the LDA+U method is required for a better description of the ground-state. The essential change in the DOS on using LDA+U is an ultraviolet shift in occupied down states. Thus it will not change our conclusions, but the laser frequency will need to be tuned based on this new LDA+U DOS.

\begin{figure}[t]
\centerline{\includegraphics[trim={0 0.5cm 0 1.6cm},clip,width=\columnwidth,angle=-0]{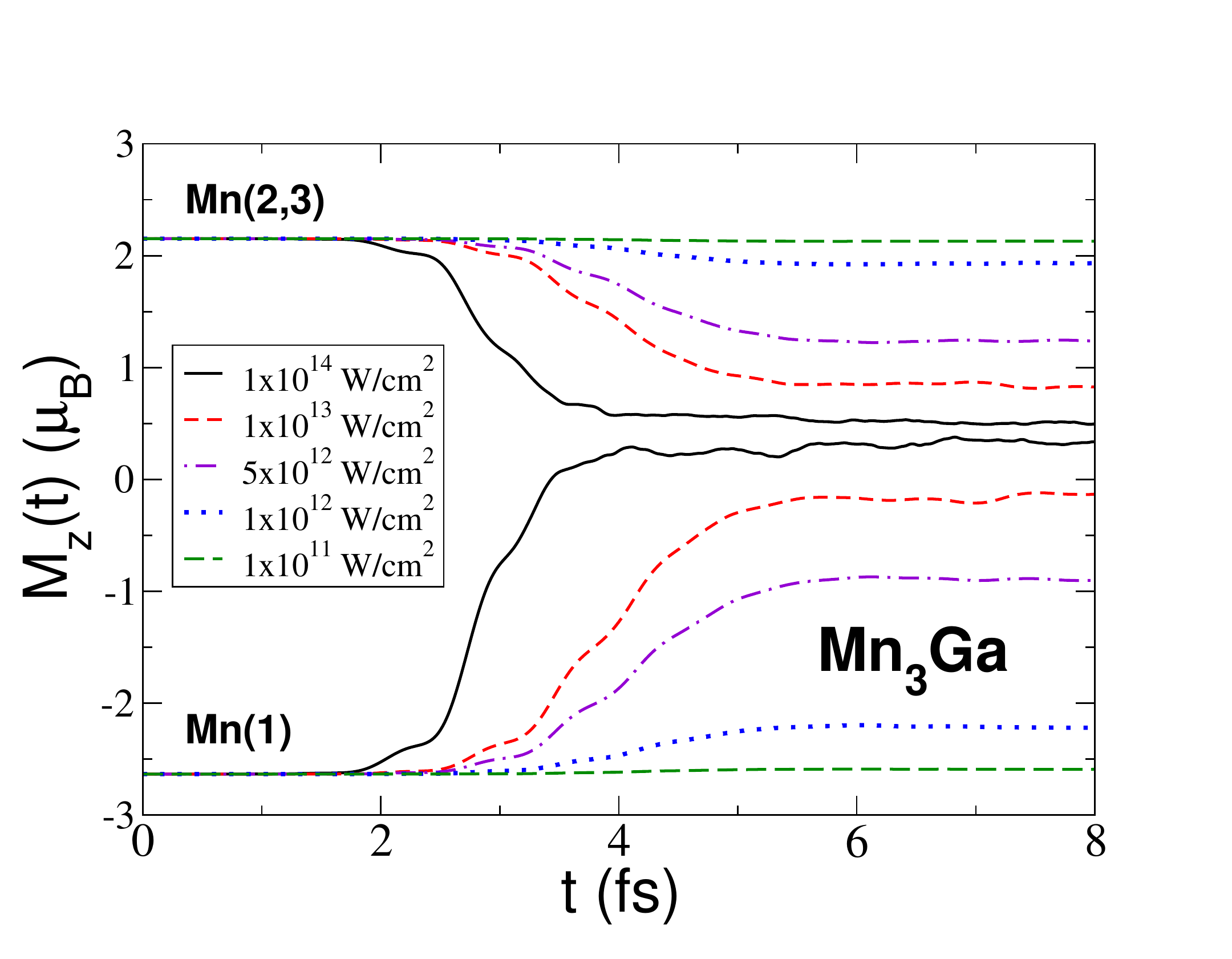}}
\caption{The dynamics of the local moment on the Mn(1) (lower curves) and Mn(2,3) (upper curves) for several different laser intensities.}
\label{f:Imn3ga}
\end{figure}

At this point it is important to note that (a) the intensity of the laser pulses used so far in this work are higher than commonly used in experiments and (b) the DOS interpretation relies on the assumption that the applied laser is a small perturbation and one can rely on ground-state DOS for analysis. However, the applied laser pulses in the present work are very intense and this assumption is not necessarily true. To investigate whether our conclusions remain applicable for significantly weaker pulses, as used in experiments, we have studied the intensity dependence of the spin dynamics in Mn$_3$Ga. In Fig. \ref{f:Imn3ga} the Mn(1) and Mn(2,3) local moments are plotted for several different laser intensities. We find that even reducing the intensity by two orders of magnitude to $1\times 10^{12}$ W/cm$^2$, still results in the same physics and a moment loss of $15\%$.

\begin{figure}[t]
\centerline{\includegraphics[width=\columnwidth,angle=-0]{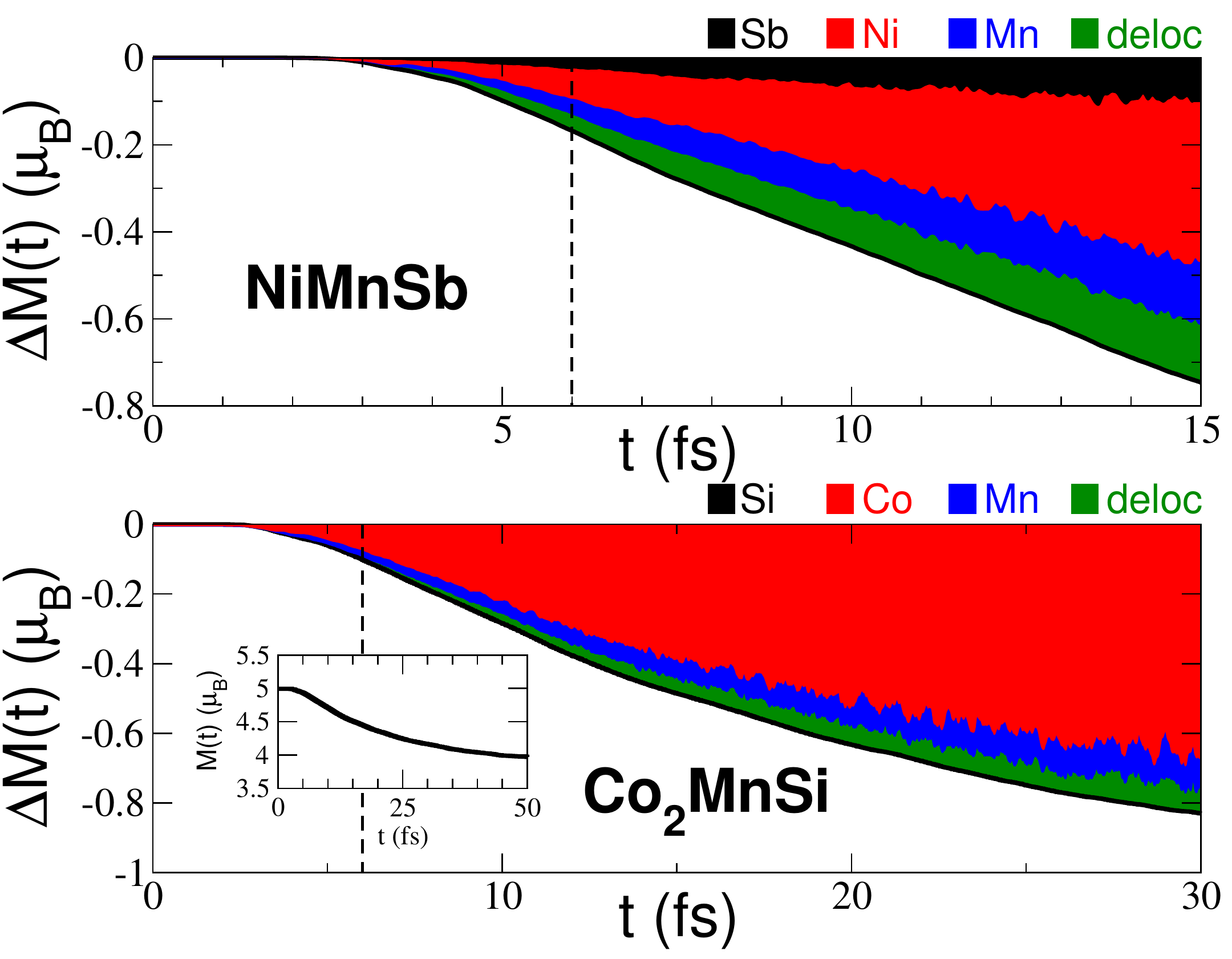}}
\caption{The longer time dynamics of the global magnetic moment for NiMnSb (upper panel) and Co$_2$MnSi (lower panel). The global moment is decomposed into the change in the local moment relative to a calculation without SOC, approximately representing where this demagnetization takes place. 
}
\label{f:MT20fs}
\end{figure}

The pulse length used in the present work is shorter than commonly used in experiments. This is intentional on our part to disentangle the physics of magnetization dynamics cause purely due to inter lattice spin-current or the one mediated by spin-orbit effect, as in Ref. \onlinecite{KDES15}. One can clearly see these two processes mixing together if one looks at the magnetization dynamics for longer times.  In Fig. \ref{f:MT20fs} such long time behavior of the global magnetic moment of NiMnSb and Co$_2$FeSi is shown. If the spin-orbit term is neglected, the spin transfer dynamics is still observed (in fact simulations with or without SOC both give the same dynamics for the first $6$ fs), however there is no loss in the global moment. On inclusion of spin-orbit coupling a global loss in moment is observed in both materials and the difference between the two calculations (with and without SOC), gives the amount of moment lost purely due to SO effects. 
In Fig. \ref{f:MT20fs}, we show the loss in global moment decomposed into contributions from each atom. In the case of NiMnSb, we see that it is primarily the Ni moment that undergoes demagnetization. Similarly, for Co$_2$FeSi major contribution to the demagnetization comes from the Co atoms. 
Even though the electrons are optically excited, the moment on Co atoms up to $6$ fs is basically unchanged from the ground-state value. The spin-orbit coupling term then acts on this optically excited Co sublattice to demagnetize it. Mn$_3$Ga was eliminated from this discussion because we do not see any strong demagnetization of the global moment for the first $20$ fs. 

Thus we have seen two possible mechanisms for changing the local moment in Heulser compounds. Firstly optical excitation can lead to inter sub lattice flow of spin current. This flow of spins may increase or decrease the local moment. The timescale for this process is  extremely short (a few femtoseconds). Secondly, spin-orbit mediated demagnetization may take place on the now out-of-equilibrium state. By considering very short laser pulses in our simulations, we could clearly distinguish these two process, however for longer duration pulses, the two effects will compete with each other. 

We have demonstrated that experimental techniques that are sufficiency sensitive to the local moment on individual atoms, e.g. the Ni atom in NiMnSb, will find a rich world of ultrafast spin dynamics to explore. This world contains multiple ways to manipulate the local moments and thus make the eventual technological transfer of such physical processes all the more viable. Similarly we have seen that the Heusler family of materials offers a range of different behaviors under the influence of ultrashort laser pulses that demonstrates the flexibility it offers. Understanding and controlling the ultrafast spin dynamics of complex materials will be of key importance to the future development and design of Spintronic devices. We show that despite the strong external fields, which knock the system out of equilibrium, in certain cases such a material design is possible from a simple ground state calculation because most of the physics of spin dynamics is controlled by the flow of spin current which is governed by availability of states above the Fermi level in the ground-state.

{\bf Acknowledgements}
We thank Dr. Kevin Kreiger for useful discussions and Dr. Yu-ichiro Matsushita (U. of Toyko) for early calculations on similar Heusler materials. We also thank Dr. Daniel Steil and Prof. Stefan Matthias (both U. of Goettingen) for helpful discussions on their experimental work.

\bibliography{demag}

\end{document}